\documentclass{jfm}
\usepackage[utf8]{inputenc}
\usepackage{graphicx}
\usepackage{natbib}
\usepackage{amsmath,amssymb,amsfonts,amssymb}
\usepackage{subfig}
\usepackage{setspace}
\usepackage{float}
\usepackage{microtype}
\usepackage[normalem]{ulem}
\usepackage{ragged2e}
\usepackage{xcolor}
\usepackage{tabularx}

\DeclareCaptionJustification{justified}{\justifying}
\usepackage{bm}
\usepackage{hyperref}
\hypersetup{
	hyperindex,
	breaklinks,
	colorlinks=true,
	linkcolor=blue,
	citecolor=magenta,
	bookmarks=true,
	bookmarksopen=true,
	bookmarksopenlevel=2,
	pdfstartpage={1},
	pdfstartview={FitH},
	pdfview={FitH 0},
	pdfauthor={M.-A. Nikolaidis and P. J. Ioannou},
	pdftitle={}}

\usepackage{ifthen}

\def\P{{\bf P}}

\def\P{{\bf P}}

\newcommand\redsout{\bgroup\markoverwith{\textcolor{red}{\rule[0.5ex]{2pt}{0.4pt}}}\ULon}

\newcommand{\be}{\begin{equation}}
\newcommand{\ee}{\end{equation}}
\newcommand{\bdm}{\begin{equation*}}
\newcommand{\edm}{\end{equation*}}
\newcommand{\bea}{\begin{eqnarray}}
\newcommand{\eea}{\end{eqnarray}}

\newcommand{\partialf}[2]
{
 \ifthenelse{\equal{#1}{}}{\frac{\partial}{\partial #2}}{\frac{\partial #1}{\partial #2}}
}

\newcommand{\<}{\left\langle}
\renewcommand{\>}{\right\rangle}

\newcommand{\Del}{\Delta}

\renewcommand{\v}{v}

\newcounter{saveeqn}%

\definecolor{mygreen}{rgb}{0,0.6,0}
\definecolor{mygray}{rgb}{0.5,0.5,0.5}
\definecolor{mymauve}{rgb}{0.58,0,0.82}

\def\bt{\tilde{\beta}}

\def\st{\sin{\vartheta}}

\def\nablav{\bm\nabla}

\newcommand{\defn}{\ensuremath{\stackrel{\mathrm{def}}{=}}}
\renewcommand{\equiv}{\defn}

\renewcommand{\u}{\mathbf{u}}
\renewcommand{\v}{\mathbf{v}}

\shorttitle{Synchronization of passive subspaces in turbulent Couette flows}
\shortauthor{M.-A. Nikolaidis and P. J. Ioannou}

\title{Synchronization of   Low Reynolds Number Plane Couette Turbulence}

\author{Marios-Andreas~Nikolaidis\aff{1}\corresp{\email{mnikolaidis@phys.uoa.gr}}  \and  Petros J. Ioannou\aff{1,2} }
\affiliation{\aff{1}Department of Physics, National and Kapodistrian University of Athens, Athens, Greece
\aff{2}Department of Earth and Planetary Sciences, Harvard University, Cambridge, U.S.A.}

\begin{document}

\maketitle

\date{\today}

\begin{abstract}

We demonstrate that a separation of the velocity field in large and small scales according to a streamwise Fourier decomposition identifies subspaces with stable Lyapunov exponents and 
allows the dynamics to exhibit properties of an inertial manifold, such as the 
synchronization of the small scales in simulations sharing the same large scales or equivalently the decay of all small scale flow
states to the state uniquely determined from the large scale flow. 
This behaviour occurs for deviations with streamwise wavelength smaller than $130$ wall units which was shown in earlier studies to correspond to the streamwise spectral peak of the cross-flow velocity components of the top Lyapunov vector of the turbulent flow.  

 \end{abstract}


\maketitle

\section{Introduction}

The dynamics governing turbulent flows are associated with a large number of degrees of freedom required to describe the velocity field in a given domain. 
In the framework of Kolmogorov's famous $K41$ theory \citep{Kolmogorov-1941} for homogeneous isotropic turbulence, we can obtain an estimate for this number from the ratio of the sizes between the largest and smallest length scales of the flow which is equal to the viscous Reynolds number ($R$) raised to a power of $9/4$.  
Even though the degrees of freedom grow rapidly as $R$ increases, the accuracy of this number relies upon assumptions in $K41$ which are generally not satisfied in the inhomogeneous setting of wall-bounded flows.
A dynamical systems approach, on the other hand, proposes the study of a potentially strange attactor underlying the turbulent state and associates the degrees of freedom with the dimension of this attractor \citep{Keefe-etal-1992} which is still large but more tractable.
The idea that not all degrees of freedom are equally important is certainly not a new concept,
since 
it is believed that turbulent flows are primarily driven by the large scale coherent motions which are 
actively participating in the energy extraction from external energy inputs, whereas the small scales are responsible for dissipation of energy that is transferred to them from the larger ones.
This energy transfer to  
the smaller scales is accomplished through a series of nonlinear interactions, 
which quite possibly could imply that the large scales exert considerable influence on the dynamics of the smaller scales.

\begin{figure}
\centering
\includegraphics[width=0.6\columnwidth]{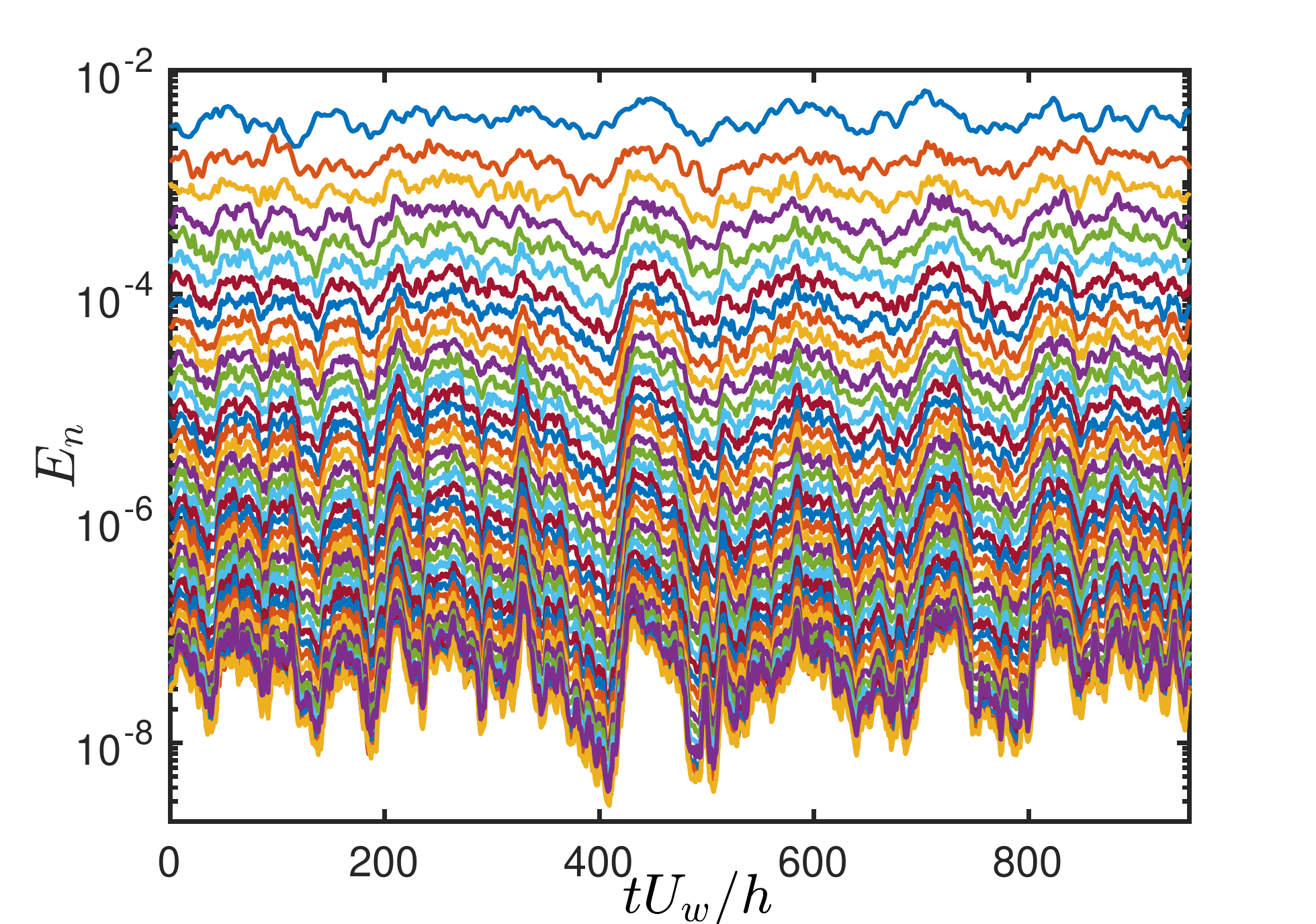}
\caption{ Time series of the energy density $E_n$ of the  39  streamwise varying components in the R1500 simulation (see Table \ref{table:geometry}). $E_n$ indicates the energy density 
of the Fourier component with streanwise wavenumber $k_x= n \alpha$ where $\alpha=2\pi/ L_x$ is the smallest non-zero wavenumber allowed in the channel
and $n=1,\cdots,39$. $E_n$  monotonically decreases with increasing wavenumber. Note that 
the higher harmonics vary in unison.}
\label{fig:1}
\end{figure}

\begin{figure}
\centering
\includegraphics[width=0.6\columnwidth]{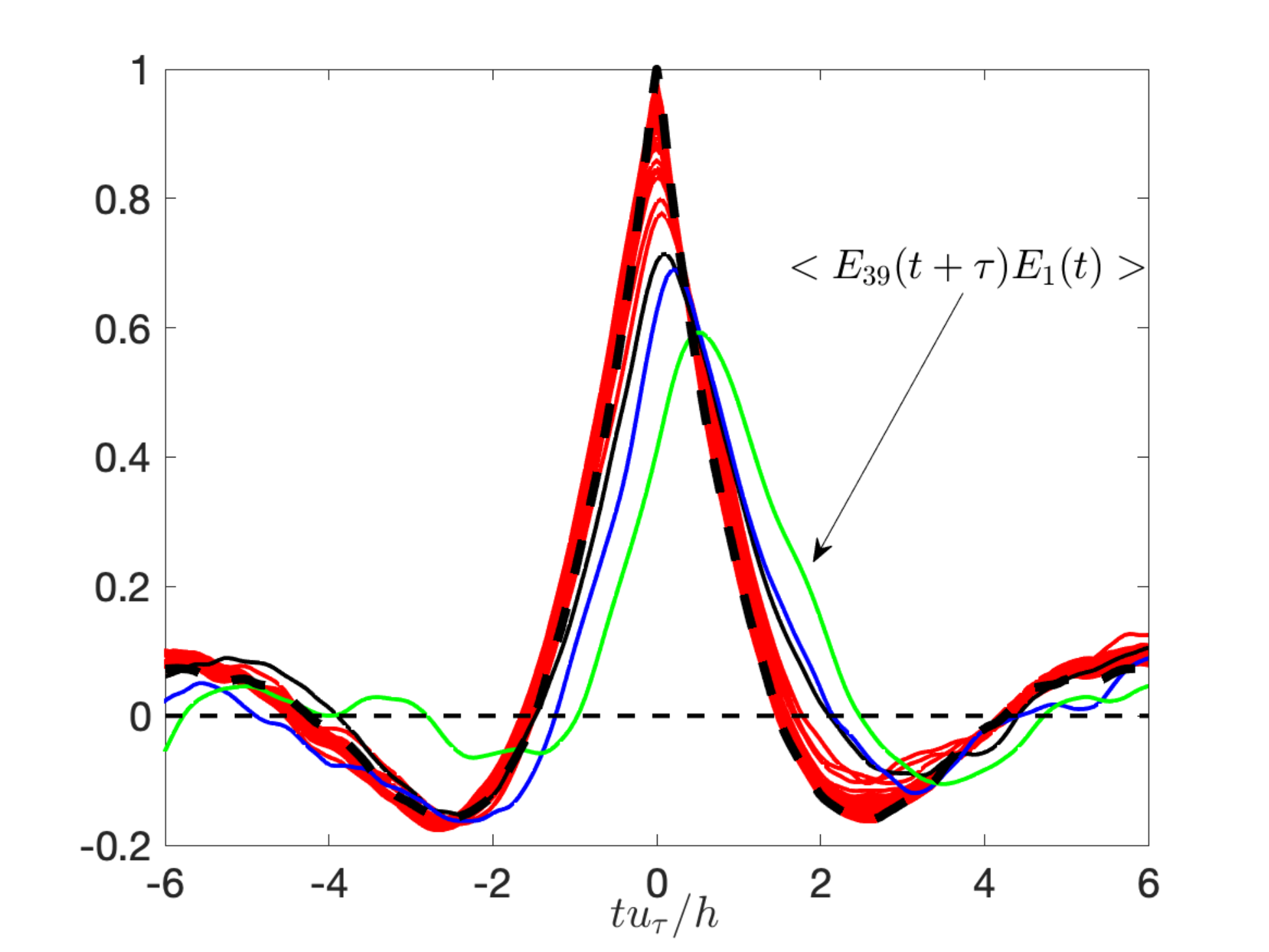}
\caption{ The normalized time lag correlation coefficient of the energy, $E_{39}$,  of the last  streamwise Fourier component retained in the DNS (with streamwise wavenumber  $39 \alpha$) with  energy $E_n$ as a function of the time lag expressed in wall units  ($u_\tau =0.0621 U_w$).
The correlation with streamwise harmonics $n=1,2,3$ is indicated with a different color.  The lag autocorrelation $<E_{39}(t+\tau) E_{39}(t) >$ is indicated with the thick dash line. The energies of  streamwise harmonics  $E_n$ with $n \ge 4$ are strongly correlated with $E_{39}$.
The case shown is for the time series shown in Fig.  \ref{fig:1}.}  
\label{fig:corrEkx}
\end{figure}

When studied in the physical space, the various scales are defined by the size of the eddies appearing in the flow.
Alternatively, we can perform a Fourier transformation of the velocity field and obtain components with designated wavelengths representative of the respective scales which allow us to formulate a theory showing that the energy extracting properties of the wavelengths correspond to their counterparts in the physical space. 
Recently, it was also shown that a self-sustained state of turbulence  in  wall-bounded flows is obtained even when the interactions
among scales in  the Navier-Stokes equations are restricted to the mere interaction between the streamwise-averaged velocity of the flow with
a single large scale streamwise  Fourier component of the flow \citep{Farrell-Ioannou-2012,Constantinou-etal-Madrid-2014,Thomas-etal-2015}.
Furthermore, curiously, it was observed in various DNS simulations of wall-bounded turbulence
that the energy of these higher order streamwise components  seems to be  synchronized to the larger scale motions \citep{Farrell-etal-2016-VLSM}.
An example of the time variation of the streamwise components in a turbulent flow at $R=1500$ is shown in Fig. \ref{fig:1} and in Fig. \ref{fig:corrEkx} it is evident that these energy time series are highly correlated in time.

Synchronization of chaotic systems \citep{Pecora-Carroll-1990} is commonly associated with collapsing subspaces that may exist in the dynamics, which have the property to attract all nearby trajectories onto a single  solution lying on that subspace. 
It is understood that synchronization requires two nearly identical dynamical systems, one that solves fully the equations of motion and another that when supplied with a fraction of the solution from the first can recover the remaining state variables.
The separation of the dynamical state into two components, as scetched above, implies the existence of an active component that is defining the state and a passive component which is determined uniquely by the active component. 
In isotropic turbulence, synchronization phenomena have been employed to reproduce the velocity field at the dissipative lengthscales of the flow when the energy-extracting and inertial lengthscales have been determined \citep{Yoshida-etal-2005,Lalescu-etal-2013,Di-Leoni-etal-2020}. It was recently proposed that the synchronization of the intense vorticity formations at those lengthscales is driven from the vortex stretching mechanism \citep{Vela-Martin-2021}.     
Additionally, the effect of the large scale motions was shown to be imprinted on the small scale statistics in turbulent wakes \citep{Thiesset-etal-2014,Alves-etal-2020}, raising questions about the meaning and universality of the Kolmogorov cascade.

The aim of this work is to investigate whether a subspace of the instantaneous streamwise spectral components of the velocity field can be recovered from the equations of motion when the remaining degrees of freedom are known. Such subspaces exist in various dissipative dynamical systems and have been denoted as the inertial manifold of the dynamics \citep{Titi-1990,Foias-etal-1993}, where any deviations from the solution decay exponentially. \cite{Lalescu-etal-2013} suggested that given the synchronization occurring in sub-Kolmogorov scales, their exclusion from a finite dimensional Navier-Stokes solution could still lead to an exact solution in the continuous space. However in this work, the energy of the synchronizable subspace is just an order of magnitude smaller than  the active subspace and it is not possible to simply neglect this part of the dynamics without affecting the accuracy of the solution.

 \section{Formulation}

In order to examine  whether the smaller scales can be slaved
to the larger scales or equivalently  whether the smaller  scales can be synchronized by  the larger
we split the flow field $\u$ into Fourier components   containing the longest streamwise scales, 
$
\u_<= \sum_{0\le n  < N}   \u_{n } e^{i n \alpha x}\ 
$,
where $k_n= n \alpha$, with  $\alpha = 2 \pi/ L_x$, is the wavenumber of the  $n$-th  streamwise harmonic in the periodic channel with streamwise length $L_x$, and to the  smaller scales 
$
\u_{\ge N}= \sum_{n  \ge N}   \u_{n } e^{i n \alpha x}
$,
with $\u = \u_< +\u_>$.

The equations governing these flow fields are obtained by projecting the  incompressible  Navier-Stokes equation
on the corresponding Fourier subspaces (cf.  \citet{Frisch-1995}). The large scales are governed by the Navier-Stokes 
 \bea
\partial_t \u_< = & -P_< \left ( (\u_<+\u_>) \cdot \nablav  (\u_<+\u_>)  - R^{-1} \Del \u_< \right )\ ,  
\label{eq:NS1} 
\eea
where $P_<$ is the Leray projection on the space spanned by streamwise harmonics $0,\cdots,N-1$, 
coupled to the Navier-Stokes for the smaller  scales
\bea
\partial_t \u_> = & -P_> \left ( (\u_<+\u_>) \cdot \nablav  (\u_<+\u_>) - R^{-1} \Del \u_> \right )\ ,  
\label{eq:NS2} 
\eea 
where $P_>$ is the Leray projection on the space spanned by the $n \ge N$ streamwise harmonics.

\begin{figure}
\centering
\includegraphics[width=0.6\columnwidth]{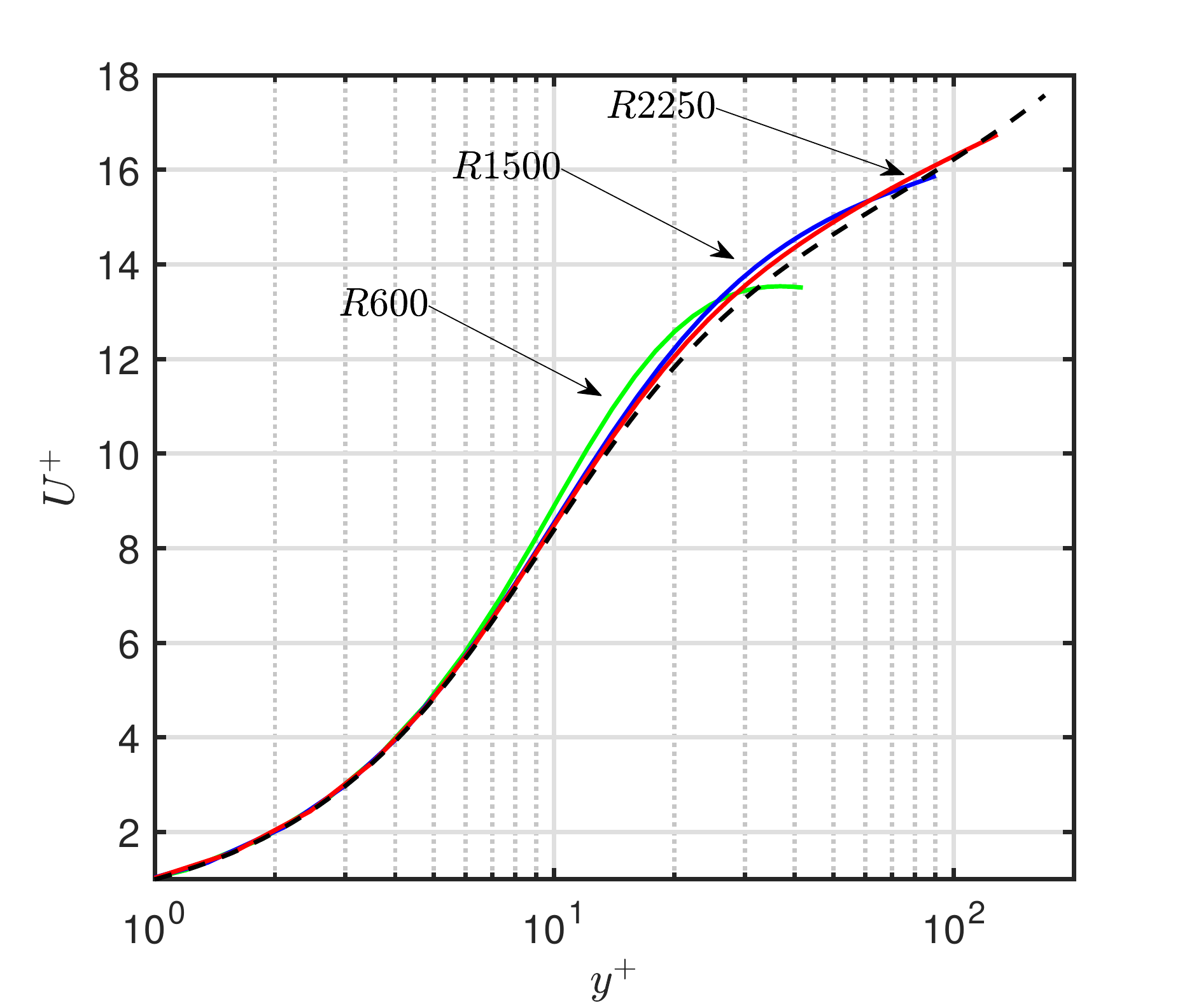}
\caption{ Streamwise, spanwise and time average of the streamwise velocity component in viscous wall units for simulations in table \ref{table:geometry}. The dashed line indicates the mean velocity profile obtained in \cite{Pirozzoli-etal-2014} for a larger plane Couette channel at $R_{\tau}=171$. }
\label{fig:m_prof}
\end{figure}

\begin{center}
\begin{table}
\caption{\label{table:geometry}Simulation parameters.  
The channel  size in the streamwise, wall-normal and spanwise  directions is $[L_x,L_y,L_z]/h=[1.75\pi,2,1.2\pi]$, where $h$ is  the half-width.   
Lengths in
wall-units are indicated by  $[L_x^+,L_y^+,L_z^+]$.
$N_x$, $N_z$ are the number of Fourier components after dealiasing and $N_y$ is the number of Chebyshev components.
$R_{\tau} =  h u_{\tau} / \nu$ 
is the Reynolds number of the simulation based on the friction velocity $u_{\tau}$ and $R = h U_w / \nu$ the bulk velocity Reynolds number according to the velocity at the wall $U_w$, the viscosity $\nu$ and the channel half-width $h$. 
$\Delta x^+$ and $ \Delta z^+$ denote the average streamwise and spanwise grid spacing in wall units.}
\centering\vspace{.8em}
\begin{tabular}{@{}*{7}{c}}
\break
 Abbreviation  &$[L_x^+,L_z^+,L_y^+]$&$N_x\times N_z\times N_y$& $R_\tau$& $R$ & $\Delta x ^+$ & $\Delta z ^+$ \\
 R600   & $[245\;,168,\;\;\;89]$&$35\times 35\times 55$&$44.5$& $600$ & $7.07$ & $4.84$ \\
 R1500   & $[512\;,351,\;181]$&$79\times 79\times 75$&$93.1$& $1500$ & $6.47$ & $4.44$ \\
 R2250  & $[744\;,510,\;271]$&$ 87\times 91\times 83$&$135.4$& $2250$ & $8.43$ & $5.53$ \\
\end{tabular}
\end{table}
\end{center}

\begin{figure}
\centering
\includegraphics[width=0.8\columnwidth]{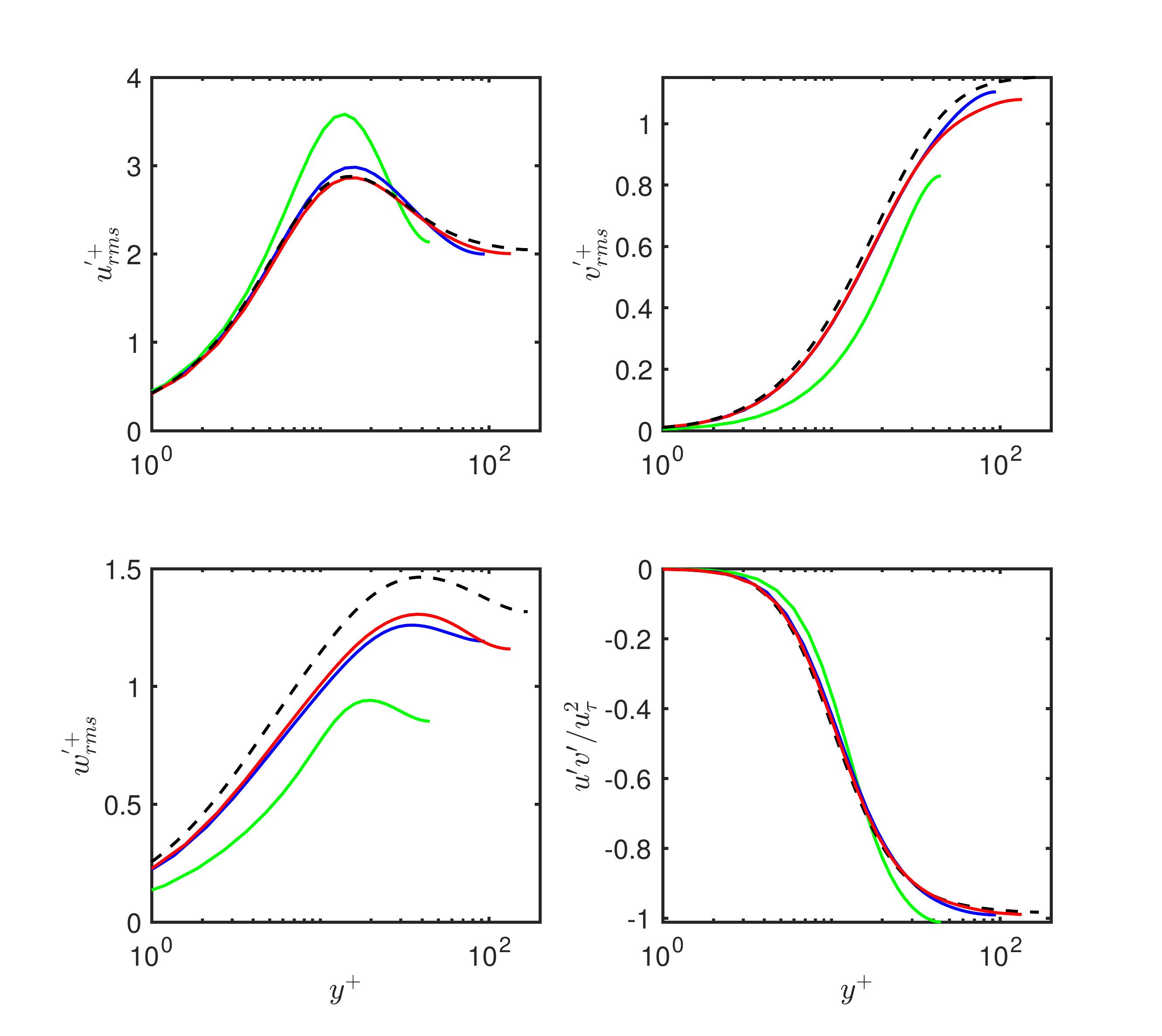}
\caption{ One-point statistics for the simulations R600 (green), R1500 (blue) and R2250 (red) in Table \ref{table:geometry}. (a,b,c) Root mean square (r.m.s.) profiles for the velocity fluctuation components $(u',v',w')=\u-U(y)$ in viscous units. 
$U$ is the streamwise, spanwise and time average of the streamwise velocity component.
(d) Streamwise, spanwise and time average profile of tangential Reynolds stress $u'v'$ in viscous units. 
The dashed line in each graph indicates the same statistics obtained in \cite{Pirozzoli-etal-2014} at $R_{\tau}=171$.}
\label{fig:m_rms}
\end{figure}

Synchronization of the smaller scales to the larger scales is achieved when an arbitrary smaller scale flow field, $\v_>$,  governed by 
the Navier-Stokes equation
\bea
\partial_t \v_> = & -P_> \left ( (\u_<+\v_>) \cdot \nablav  (\u_<+\v_>) - R^{-1} \Del \v_> \right )\  
\label{eq:NSs} 
\eea 
  in the smaller Fourier scale subspace  with prescribed  large scale field $\u_<$  satisfying Eqns.  \eqref{eq:NS1} and \eqref{eq:NS2},
converges to  $\u_>$, i.e. synchronization with the DNS realization is achieved when 
$\lim_{t\to \infty} || \v_>- \u_> || = 0$. We choose as metric $|| \cdot ||$  the square root of the kinetic energy density of the flow.

When the top Lyapunov exponent of the linearized equation \eqref{eq:NS2} governing the smaller scales 
about  a prescribed $\u_<$ and $\u_>$  obtained from the NS \eqref{eq:NS1} and \eqref{eq:NS2}
is negative then an arbitrary field  $\v_>$ will synchronize to $\u_>$, at least when the $\v_>$ is adequately close to $\u_>$ and is within the attractor basin of $\u_<+\u_>$.
This leading  Lyapunov exponent is obtained by calculating the exponential rate of evolution of the perturbation field $\u_>'$
governed by  equation \eqref{eq:NS2} linearized about the DNS solution $\u_<$, $\u_>$:
 \bea
\partial_t \u_>' = & -P_> \left ( (\u_<+\u_>) \cdot \nablav  \u_>' + \u_>' \cdot \nablav  (\u_<+\u_>)  - R^{-1} \Del \u_>' \right )\ ,  
\label{eq:NSp} 
\eea 
These linearized equations, which are referred to as the variational equations,   determine the asymptotic stability of  the time-dependent flow $\u_<+\u_>$
of  the DNS to perturbations  $\u_>'$ with streamwise
wavenumbers $k_x \ge N \alpha$. 
 We  denote with $\lambda_N$ the top Lyapunov exponent of \eqref{eq:NSp} for this  truncation, which is  defined as:
\bea
\lambda_N =  \overline{\lim_{t \to \infty}}\frac{1}{t} \log \left (||\u_>'(t)|| \right ).
\label{eq:lamdaN}
\eea 
The top Lyapunov exponent that determines the sensitivity of the whole turbulent flow field  is $\lambda_0$ and its properties are discussed for channel flows  in \cite{Keefe-etal-1992,Nikitin-2008,Nikitin-2018}.

\section{Determination of the Lyapunov exponents of the turbulent flow to streamwise restricted perturbations }

We consider DNS of  turbulent Couette flows at $R=600$ that sustains turbulence with $R_\tau = 45$, at $R=1500$ with $R_\tau = 93$ and at $R=2250$ with $R_{\tau}=135$ 
and calculate the top Lyapunov exponents $\lambda_N$  of \eqref{eq:NSp} for various truncations $N$. Our simulations are performed with an in-house developed direct numerical simulation MATLAB code with GPU acceleration that solves the NS equations in the form proposed by \cite{Kim-etal-1987}, and employs Chebyshev discretization on the wall-normal direction and a finite difference grid on the streamwise and spanwise directions which are treated pseudo-spectrally and are dealiased following the $2/3-$rule. Time-stepping is accomplished with a Crank-Nicolson/3rd order Runge-Kutta scheme for the viscous and advective terms respectively. Parameters of the simulations are summarized in Table \ref{table:geometry}. The one-point flow statistics of the simulations and their comparison with the literature are plotted in Fig. \ref{fig:m_prof} and \ref{fig:m_rms}.

\begin{figure}
\centering
\includegraphics[width=0.8\columnwidth]{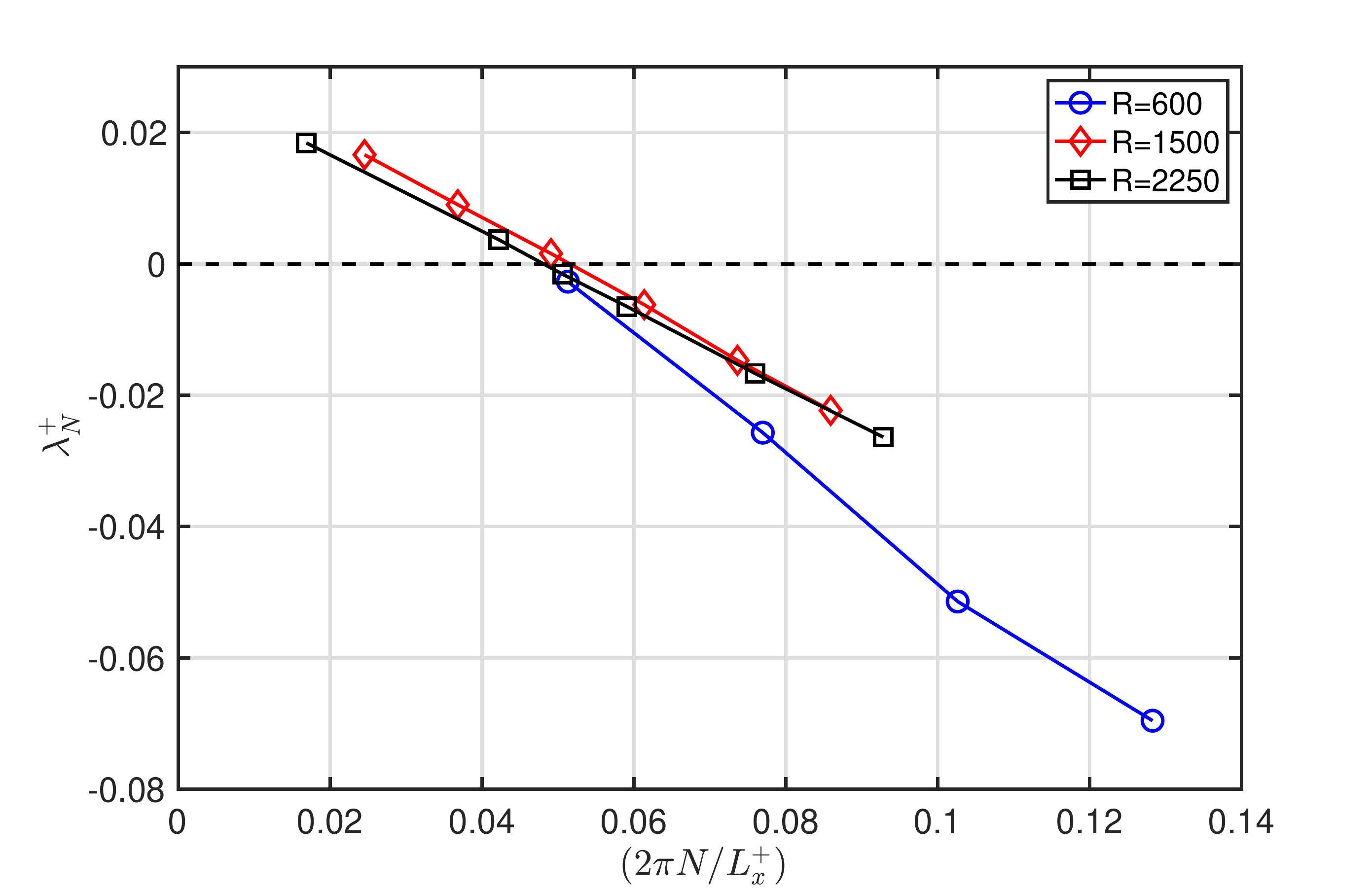}
\caption{ The top  Lyapunov exponents, $\lambda^+_N$, of the variational equations  Eq. \eqref{eq:NSp} as a function
of the wavenumber $2 \pi N / L_x^+$ of the gravest scale included in  $\u_>$, plotted in viscous time units $t^+=(h u^2_{\tau}/ \nu)$. 
For $R=600$ we plot $\lambda_N$ for 
$N=2,3,4,5$  (blue circles), for $R=1500$ we plot $N=2,3,4,5,6,7$  (red diamonds),  and $N=2,5,6,7,9,11$ for $R=2250$ (black square).  
This figure indicates  that at $R=600$ the critical streamwise harmonic  for synchronization is $N_c=2$ 
and the $\u_<$ flow components  with streamwise harmonics $n=0,1$
synchronizes  the $\u_>$ flow. This is verified in the simulation shown in Fig. \ref{fig:sync600}. 
At $R=1500$ synchronization occurs for $N_c=5$ when the $\u_<$ field
includes streamwise harmonics $n=0,1,2,3,4$, as shown in Fig. \ref{fig:sync1500}. At $R=2250$,  $N_c=6$ and  the 
$\u_<$ field  comprised of $n=0,1,2,3,4,5$  synchronizes the $\u_>$ flow.}
\label{fig:lambdaN}
\end{figure}

The numerical integration of eq. \eqref{eq:NSp} for the determination of the top Lyapunov exponent is initialized from a randomly generated, divergence-free initial condition that lies on the specified subspace, which is kept infinitesimal by normalization at the end of each time-step. The fields $\u_<$ and $\u_>$ appearing in \eqref{eq:NSp} are determined simultaneously by the DNS of \eqref{eq:NS1} and \eqref{eq:NS2}. An instantaneous growth rate, $\lambda(t)$, is then calculated from the ratio of $||\u_>'||$ at time $t$ and at the adjacent time-step $t+\Delta t$ before normalization,
\bea
\lambda(t)= \frac{1}{\Delta t} \log \left (\frac{||\u_>'(t+\Delta t)||}{ ||\u_>'(t)||} \right ).
\label{eq:lamda}
\eea 
The $\lambda(t)$'s are highly variable in time, but in the long-term their average value converges 
to the   Lyapunov exponent $\lambda_N$, for each choice of $N$. 
The flow states associated with the Lyapunov exponents have non-negligible spectral coefficients in all included streamwise  wavenumbers (i.e. $k_x \ge N \alpha$).

The resulting top Lyapunov exponents are plotted in Fig. \ref{fig:lambdaN} as a function of the order of truncation $N$.
As $N$ increases  
the top Lyapunov exponent monotonically decreases  and eventually   becomes negative at $N_c$, indicating that subspaces with $N\ge N_c$  are in principle synchronizable. 
A negative Lyapunov exponent is 
located at different $N_c$ for the three Reynolds numbers considered.  
 When the  wavelengths and the growth rates   are scaled with the corresponding friction velocities, the predicted threshold 
streamwise wavelength at which $\lambda_N$  becomes negative collapses to the value $l_x^+=130$. Moreover, the Lyapunov exponents appear to converge to a single curve as the Reynolds number increases and the size of the channel becomes unimportant \citep{Inubushi-etal-2015,Nikitin-2018}.  
The same type of scaling was shown by \citep{Nikitin-2008} to produce a Lyapunov exponent of  $\lambda_0^+=0.021$ for the total flow in channels and pipe flows with $R_{\tau}$ between $140$ and $320$.

\section{Synchronization experiments}

\subsection{Synchronization of the passive subspace}

\begin{figure}
\centering
\includegraphics[width=0.8\columnwidth]{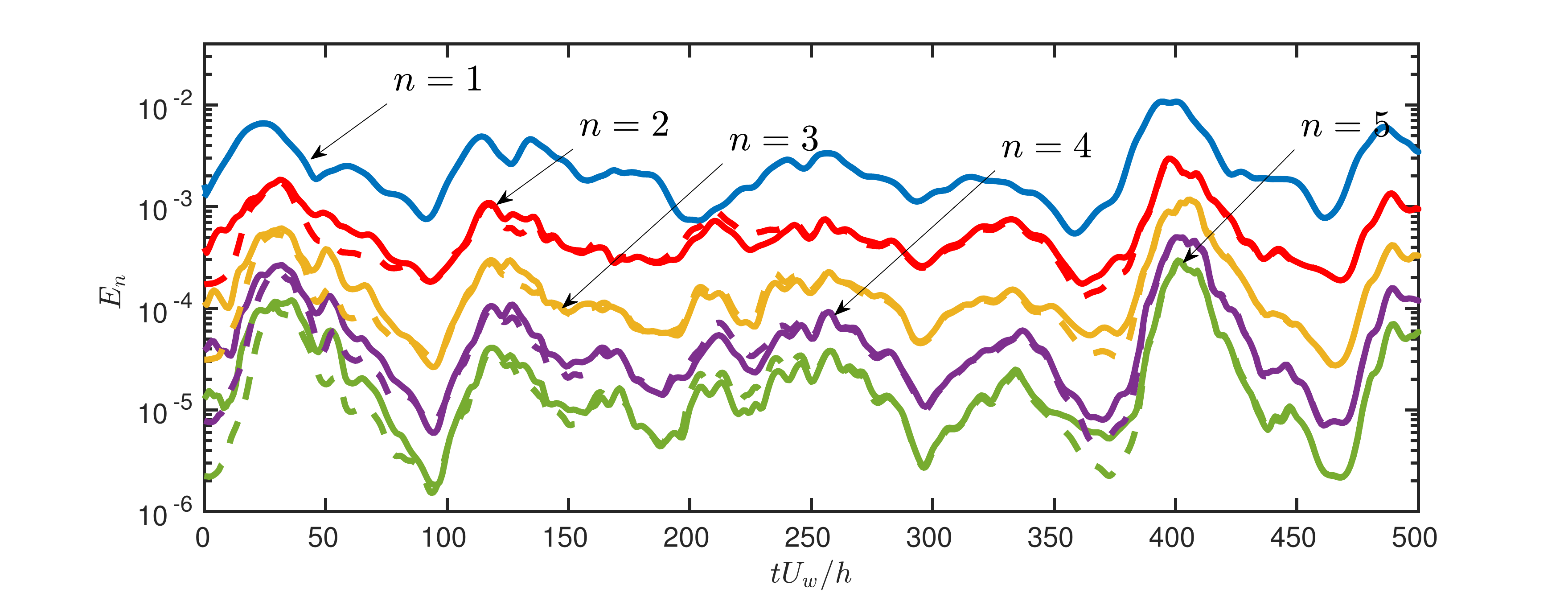}
\caption{ Time series of the energy density $E_n$ of the  first 5  streamwise varying components in the R600 simulation (solid lines).
In this experiment the primary field $\u_<$ with streamwise harmonics $n=0$ (not shown) and $n=1$ synchronizes the $N=2$ velocity field $\v_>$ (dashed lines) to the $\u_>$  of the DNS. 
The exponential rate of convergence towards synchronization is  $0.01 U_w/h$ (cf.  Fig. \ref{fig:error1500}) consistent with the decay estimated from the Lyapunov exponent $\lambda_2$ of Fig. \ref{fig:lambdaN}.}
\label{fig:sync600}
\end{figure}

\begin{figure}
\centering
\includegraphics[width=0.8\columnwidth]{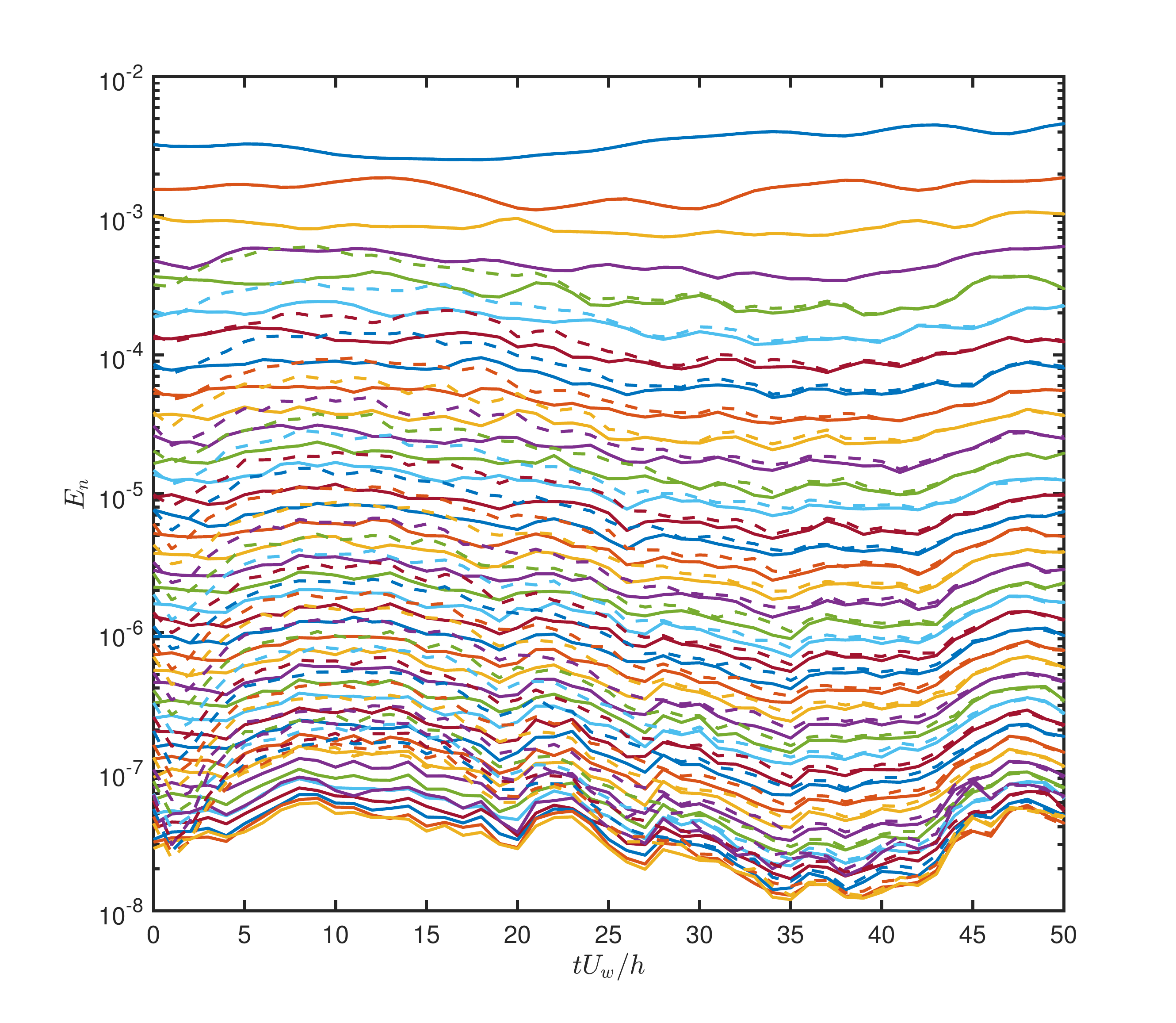}
\caption{ Time series of the energy density $E_n$  of all the  streamwise varying components of the DNS simulation of  plane Couette turbulence at 
$R=1500$ ($R_\tau= 93$) (solid lines).
With dashed lines is the energy  of the $N=5$, $\v_>$, components of the flow  coevolving with  the $\u_<$ of the DNS 
that eventually get synchronized  with the  $\u_>$  of the DNS simulation.
The exponential rate of synchronization is   $0.038 U_w/h$ (cf. Fig. \ref{fig:error1500})   consistent with the Lyapunov exponent  $\lambda_5$ of
Fig. \ref{fig:lambdaN}.}
\label{fig:sync1500}
\end{figure}

 \begin{figure}
\centering
\includegraphics[width=0.8\columnwidth]{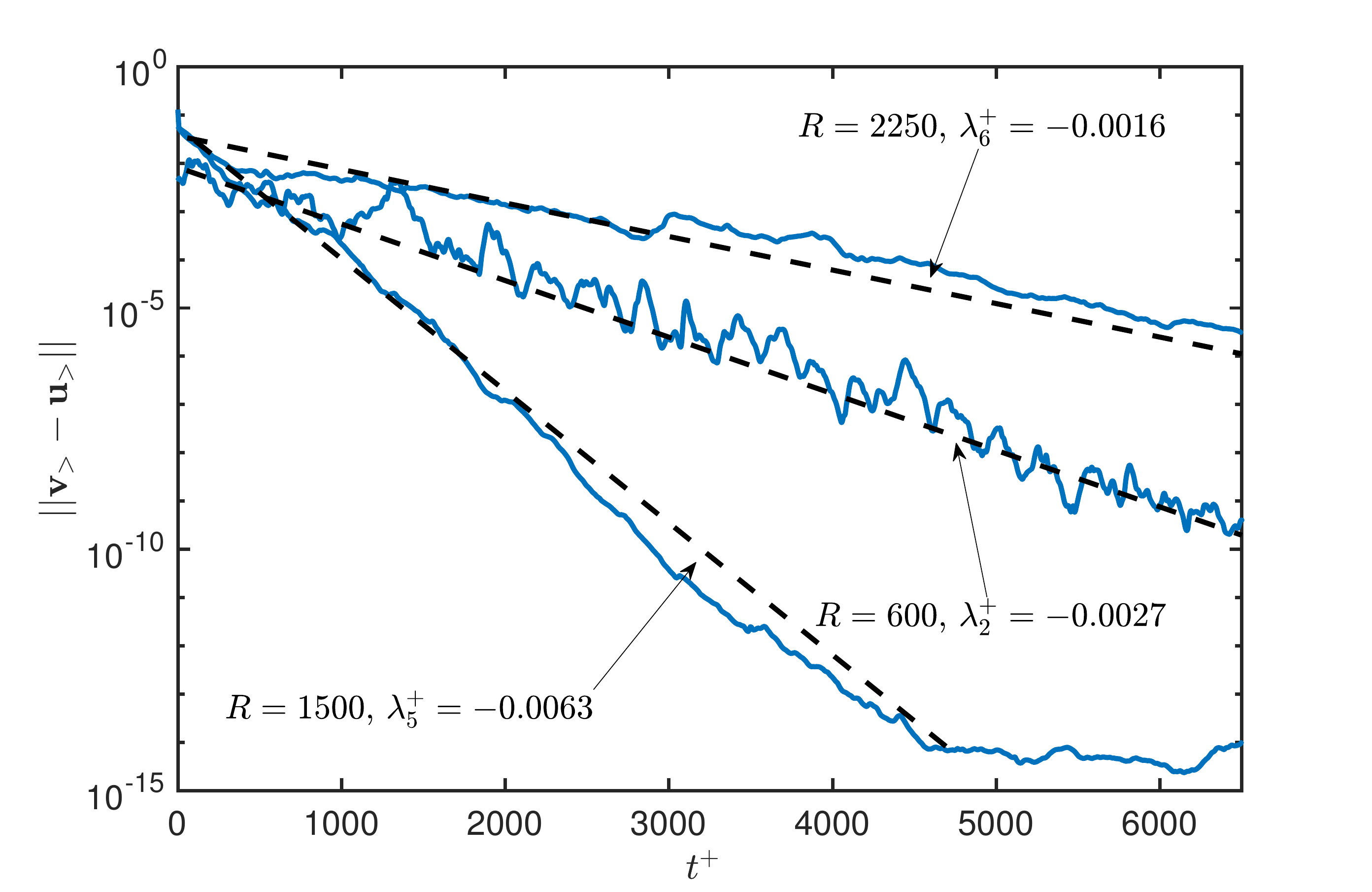}
\caption{ The approach towards synchronization of the  $\v_>$ subspace, measured by the state error norm,  $|| \v_> - \u_>||$,
 as a function of  time expressed in wall units is shown to be exponential at the rate of the Lyapunov exponent $\lambda_N^+$ (dashed). 
 Shown are the synchronization of the $N=2$ subspace at $R=600$,  of the $N=5$ subspace at $R=1500$ and 
 of the $N=6$ subspace at $R=2250$.}
\label{fig:error1500}
\end{figure} 

To perform the synchronization experiment, we couple two DNSs, one solving for the total velocity
and a second one where the components of the $\u_<$ subspace obtained from the first simulation are imposed to the second, while solving Eq. \eqref{eq:NSs} for the $\v_>$ flow field which is now initialized from an arbitrary state. 
We attempt to synchronize the subspaces $\u_>$ with  $N=N_c$ in all the cases mentioned at Table \ref{table:geometry}.
The critical truncation is predicted from the calculation of the Lyapunov exponents to be $N_c=2$ in $R600$, $N_c=5$  in $R1500$,
and $N_c=6$ in $R2250$.
 As shown in Fig. \ref{fig:sync600} for the $R600$ and in Fig. \ref{fig:sync1500} for the $R1500$, the energy density $E_n$ of the streamwise components in the $\v_>$ flow field (shown with dashed lines) converges  to the energy density of $\u_>$.
To verify that  the states $\v_>$ tend to $\u_>$, we plot the time evolution of the 
norm  of their difference $|| \v_> - \u_>||$ in Fig. \ref{fig:error1500}, which shows  that the approach towards synchronization 
is at the rate of the calculated $\lambda^+_N$  and the  error in the synchronization 
decays to numerical precision.
All streamwise components of $\v_>$  converge to $\u_>$ at the same rate.
It is somewhat counterintuitive that the $R1500$ case synchronizes earlier than the $R600$, since we 
would expect that the decay would be faster in the more viscous experiment. However, 
the results are consistent with the predicted growth rates of the Lyapunov stability analysis that are not determined solely by the viscosity.

Although the Lyapunov exponents were obtained for infinitesimal disturbances, and when negative they guarantee synchronization
of those $\v_>$ that are in an appropriately  small  neighborhood of $\u_>$,
synchronization was found to occur for states $\v_>$ that  differ from $\u_>$ by a finite amplitude.
It was also verified that even a $\v_>$ state initially set to $0$ will synchronize at the rate predicted by $\lambda_N$.
There is therefore strong indication that the attractor basin of $\u_>$ covers the whole space, and that  
the subspace $\u_>$ is uniquely determined by the subspace $\u_<$, since it attracts any $\v_>$ set as initial condition.   

\begin{figure}
\centering
\includegraphics[width=\columnwidth]{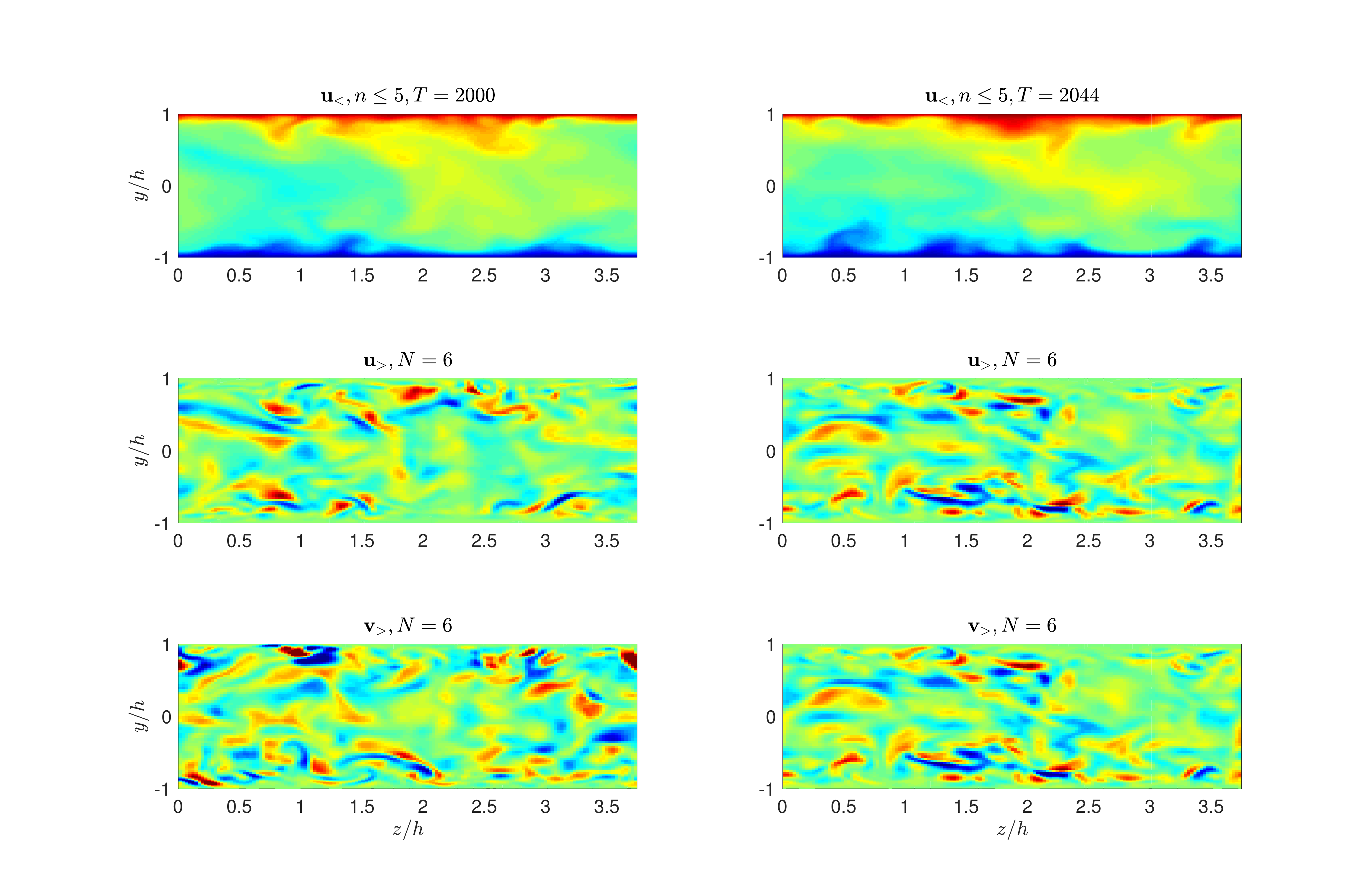}
\caption{ Contours in a $y-z$ plane cross-section of two snapshots of the streamwise velocity component of the $\u_<$, $\u_>$ and $\v_>$ flow fields
in  DNS at $R=2250$ and $N=6$. The $\u_<$ includes streamwise harmonics $n=0,\cdots,5$. 
The left panel shows the initial state of the synchronization experiment at $U_w T /h =2000$. Initially $|| \v_> - \u_>||=0.058$  while 
at $U_w T /h =2044$ the state has been almost synchronized with  $|| \v_> - \u_>||=0.007$.}
\label{fig:state5yz}
\end{figure}

\begin{figure}
\centering
\includegraphics[width=\columnwidth]{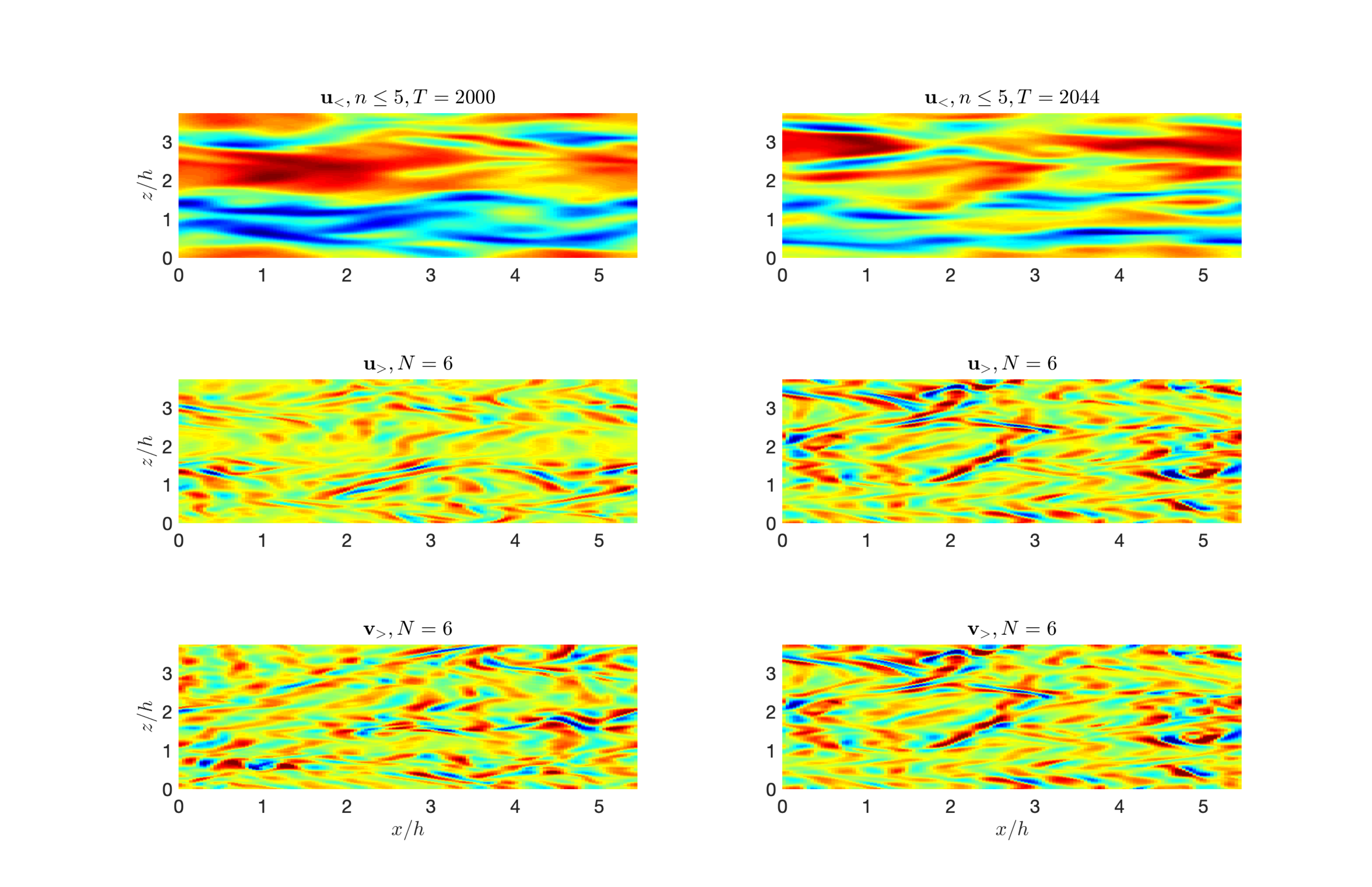}
\caption{ Contours in a $x-z$ plane cross-section at $y/h=-0.77$ of the snapshots in Fig. \ref{fig:state5yz} of the streamwise velocity component of the $\u_<$,$\u_>$ and $\v_>$ flow fields
in  DNS at $R=2250$. 
 The $\u_<$ includes streamwise harmonics $n=0,\cdots, 5$.}
\label{fig:state5xz}
\end{figure}

It is informative to  see the structure of the component of the  flow  that is synchronized.  
Indicative  velocity fields  of the $\u_<$ and $\u_>$ subspace and of the subspace $\v_>$ of the coupled dynamics at  $R2250$ with $N=6$ are shown in 
Fig. \ref{fig:state5yz} and Fig. \ref{fig:state5xz} at the initial time $U_w T/h=2000$ when an arbitrary  $\v_>$ is imposed and  at $U_w T/h=2044$
when $\v_>$ has been synchronized to $\u_>$. 
The cross-flow and spanwise ($y-z$) plane cross-section (Fig. \ref{fig:state5yz}) shows that the amplitude of the $\u_>$ component is substantial,
it reaches  a velocity of $0.15U_w$ and is more prominent in the region surrounding the buffer layer streaks. Those plots also show the streaky large scale structure of the $\u_<$ subspace 
that comprises the active part of the flow. In that sense, the $\u_>$ subspace denotes the passive part of the flow which is uniquely determined from the state of the active subspace. 

\subsection{Synchronization of single streamwise harmonics and their relation to the components sustained in RNL}

A second experiment was performed in order to identify if the lack of synchronization that occurs when $\u_>$ includes  scales with $n<N_c$, where $N_c$ is the 
critical wavenumber  for which all flow components with $n \ge N_c$  synchronize,  
can be traced to divergence caused by single streamwise components, $\v_{n}$. For this experiment, the dynamics of the single streamwise component 
$\v_{n}$ are coupled to
 all the other components of the flow $\u_{n^c}\equiv \sum_{n'\ne n}  \u_n'$  of a full DNS, according to: 
 \bea
\partial_t \v_n = & -P_n \left ( (\u_{n^c}+\v_n) \cdot \nablav  (\u_{n^c}+\v_n) - R^{-1} \Del \v_n \right )~.
\label{eq:NSsi} 
\eea 
 The single component $\v_n$ is then synchronizable if it approaches with time the streamwise component $\u_n$ of the DNS.

We have verified that when $\v_{n}$ is a component of a synchronizable subspace (i.e. if $n>N_c$)  its' state rapidly converges to that of $\u_{n}$, which is unsurprising given the synchronization of the whole subspace. Interestingly though,  some $\u_{n}$ that belong to the active subspace   with
$n<N_c$ are also found to synchronize, up to but excluding the 
last component that is maintained in a restricted nonlinear (RNL) simulation of the same flow.

In RNL simulations, the streamwise constant flow $\u_0$ (corresponding to the $\u_<$ subspace) interacts quasi-linearly with the $N=1$  streamwise varying components of the
$\u_>$   flow field\footnote{More general quasi-linear models can be  formulated with   $\u_<$ not restricted  to the single $n=0$ harmonic cf.
\cite{Constantinou-etal-2016,Marston-etal-2016}.},  according to the dynamics:
 \bea
\partial_t \u_{0}  &=&  P_0 \left ( - \u_{0} \cdot \nablav  \u_{0} + R^{-1} \Del \u_{0} -  \u_> \cdot \nablav  \u_> \right )~,
\label{eq:RNL0} \\   
\partial_t \u_>  &=&  P_> \left ( - \u_{0} \cdot \nablav  \u_> - \u_>\cdot \nablav  \u_{0} + R^{-1} \Del \u_>  \right )\label{eq:RNL1} ~.
\eea 
The restriction  in the RNL system is that in Eq. \eqref{eq:RNL1} the  $\u_>$ components interact  only  with $\u_0$,  and all other  non-linear
interactions are neglected.  The restricted dynamics produce  a self-sustained realistic state of turbulence in channel 
flows  which sustains  a flow $\u_>$ with only a few nonzero streamwise 
harmonics \citep{Farrell-Ioannou-2012,Thomas-etal-2013-structure,Bretheim-etal-2015,Farrell-etal-2016-VLSM,Farrell-etal-2016-PTRSA}. 
The few  streamwise varying components that are maintained in RNL can be characterized as
the active subspace of the turbulent flow.

The active subspace of RNL is however smaller than the active  subspace of the synchronization experiments,
 which  comprises
all  streamwise components with $n< N_c$. For example,
 at $R=600$ RNL supports one streamwise varying mode, the $n=1$ streamwise component, 
 and in this case the active subspace identified by RNL coincides with the active subspace
that achieves synchronization of the rest of the flow.  At $R=1500$ and $R=2250$  RNL supports
$n=1,2,3$, while the active subspace for synchronization at $R=1500$ includes also $n=4$ as $N_c=5$, and at $R=2250$ it includes both $n=4$ and $n=5$ as $N_c=6$.

\begin{figure}
\centering
\includegraphics[width=\columnwidth]{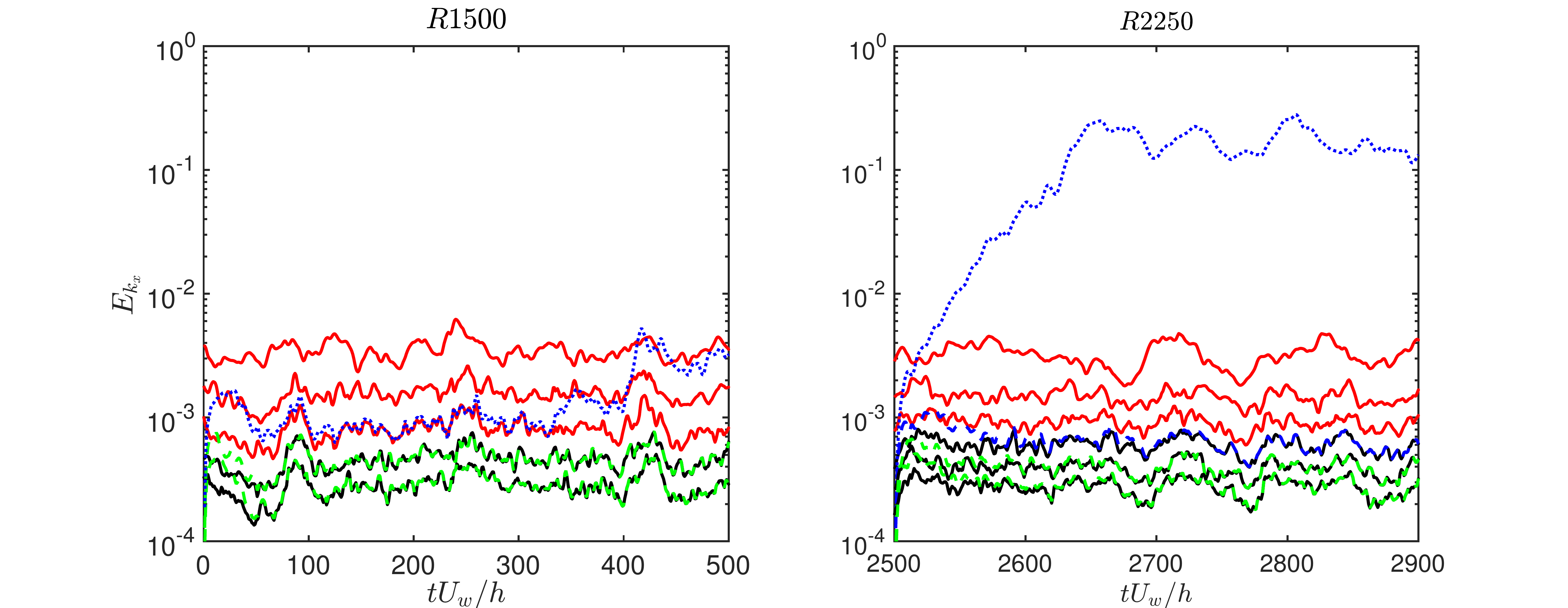}
\caption{ {
Time series of the energy of the 5 streamwise harmonics $n=1,2,3,4,5$ in a DNS at $R=1500$ (left panel) (with red are the components that are active in  RNL, with  black the  RNL inactive) and of the
6 streamwise components $n=1,2,3,4,5,6$ in a DNS at $R=2250$ (right panel)  (red - RNL active, black - RNL inactive). The energy monotonically decreases with increasing wavenumber.
Dashed or dotted lines indicate successful or failed synchronization experiments respectively, either of single components (blue) or of component pairs (green).
Left panel. The $n=3$ RNL active  component (dotted blue) fails to synchronize, while the inactive in RNL pair $n=4,5$ (dashed green) synchronizes.
Right panel. The $n=3$ RNL active component component (dotted blue) fails to synchronize, while  the $n=4$ RNL inactive component (dashed blue) synchronizes as well as the pair $n=5,6$ (dashed green).}}
\label{fig:sync2_1500}
\end{figure}

 The question arises whether the harmonics that have $n<N_c$, but are not sustained in RNL, are
synchronizable and whether the $n<N_c$ harmonics that are sustained in RNL can be synchronized.
We find that the harmonics that are sustained in RNL are not synchronizable while all other components are. 
For example, synchronization experiments of single streamwise components (or even a pair of uncoupled components) in $R1500$ (shown in Fig. \ref{fig:sync2_1500}) demonstrate that the $n=3$ streamwise component,
which is active in RNL,  does not synchronize whereas the pair  $n=4,5$ do altough $N_c=5$ in this case.
We attempted to broaden  the size of the subspace  to be synchronized  further,  but when we included the next harmonic
$n=6$ and  the subspace to be synchronized included the three harmonics, $n=4,5,6$,  it became
unstable, recovering at that level of truncation  the instability of the whole passive subspace $\u_>$  when the  $n=4$ component is included. 
This was also confirmed in the  $R2250$ experiments where again the component $n=3$, which is sustained in RNL,  was not able to synchronize,
while all other  higher harmonics were.
The significance of this result lies in the identification of the RNL subspace as the source of chaotic dynamics, as being the central subspace 
associated with the energy extraction mechanism sustaining turbulence  and through  nonlinear scattering feeding and pacing the other scales.

 \section{Energy balances of the streamwise flow components}

Energy transfers originating from the coherent large scales have been already shown to influence the 
statistics of the turbulence that operates in the small scales of inhomogeneous flows
\citep{Alves-etal-2020,Thiesset-etal-2014}.
Here we consider that the large scale coherent structure is the steamwise mean flow $\u_0$, which includes the  streak and streamwise constant rolls,
and examine the relative strength of the energy transfers to a given streamwise component  from  $\u_0$ 
and the other flow components, $\widetilde{\u}$. In this decomposition  $ \u = \u_0+\widetilde{\u}$.

\begin{figure}
\centering
\includegraphics[width=0.8\columnwidth]{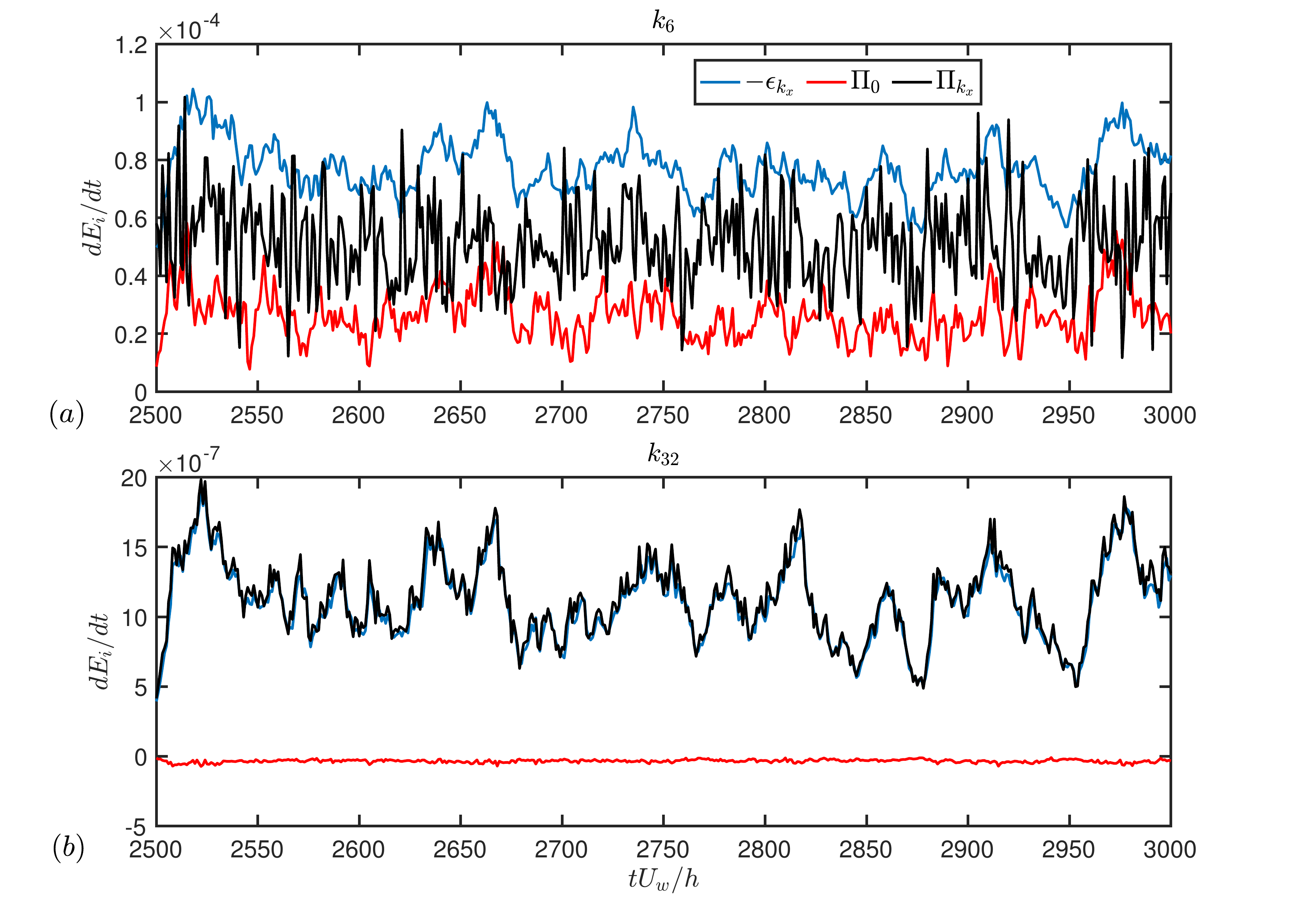}
\caption{ Time-series of the energetic terms in  \eqref{eq:Ek} of a $R2250$ simulation  for  $k_x=6\alpha$ in panel (a)  and $k_x=32\alpha$ in panel (b).
Shown are the energy transfer to $k_x$ from the mean flow $\Pi_0$ (red),  the energy dissipation rate $-\epsilon_{kx}$ (blue) and  the total 
energy transfer by the nonlinear interactions, $\Pi_{kx}$ (black).}
\label{fig:Ekx_Uw}
\end{figure}

To derive  
the energetics equation of the streamwise component with wavelength $k_x=n \alpha$, we form the inner product of the equation governing $\u_{n}$:
\bea
\partial_t \u_n = & -P_n \left ( \u \cdot \nablav  \u - R^{-1} \Del \u \right )~,
\label{eq:NSen} 
\eea 
for  $n \ge 1$, with $\u_n$ to  obtain the equation for the energetics:
\bea
\dot E_{k_x} = & \Pi_0 + \Pi_{k_x}  - \epsilon_{k_x}.  
\label{eq:Ek} 
\eea
$\dot E_{k_x}$ is the time rate of the energy of the $n$ streamwise component of the flow $ \langle ||\u_n||^2/2 \rangle $,
(angle brackets denote integration over the flow domain),
\be
\Pi_0=    - \langle \u_{n},~ P_{n} \left ( \u_0 \cdot \nablav  \widetilde{\u}  + \widetilde{\u} \cdot \nablav  \u_{0} \right )\rangle ~,
\ee
is the rate of energy transfer from $\u_0$ to  streamwise component $n$, 
 \be
 \Pi_{k_x} = - \langle \u_{n},~ P_{n} \left ( \widetilde{\u} \cdot \nablav  \widetilde{\u} \right )\rangle~,
 \ee 
 is the rate of energy transfer from the other scales to $n$ and
 \be
  \epsilon_{k_x} =   - \< \u_{n},R^{-1} \Del \u_{n}\> ~,
  \ee
is the rate of dissipation. Time series of the three terms in \eqref{eq:Ek} are plotted in Fig. \ref{fig:Ekx_Uw} for $n=6$ and $n=32$ for a $R=2250$ simulation.

\begin{figure}
\centering
\includegraphics[width=0.8\columnwidth]{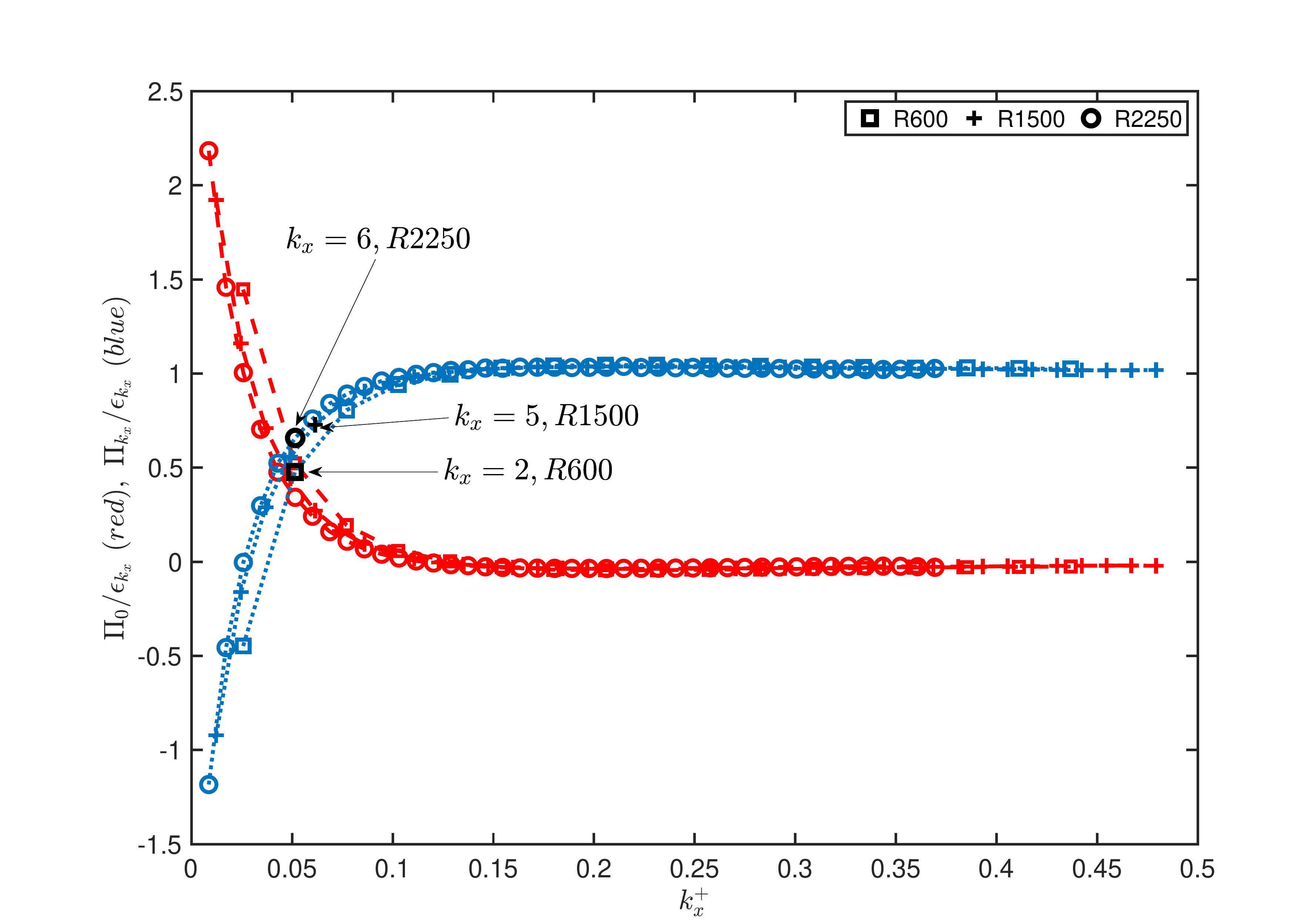}
\caption{ Comparison of the ratio of the linear $\Pi_0$ and nonlinear $\Pi_{k_x}$ transfer terms to dissipation, $\epsilon_{kx}$,  as a function of
the wavenumber, $k_x^+$  in wall units for  simulations at $R=600$, $R=1500$ and $R=2250$. 
All curves collapse to a single line with this scaling.  The streamwise scale that renders
$\u_>$ slave occurs  in the vicinity of the intersection  of the two curves, $k_{xc}^+$,  indicated
with black markers (square for $R=600$, cross for $R=1500$ and circle for $R=2250$).  }
\label{fig:energetics1500}
\end{figure}

 The terms of Eq. \eqref{eq:Ek} are evaluated for every streamwise component, and the ratios of the time-averaged quantities, $\Pi_0/ \epsilon_{k_x}$ and $\Pi_{k_x} / \epsilon_{k_x}$, are plotted as functions of the streamwise wavenumber scaled in viscous wall units, $k^+_x = n \alpha/ R_{\tau} $, in Fig. \ref{fig:energetics1500}. The scaling shows that the wavenumber $k_{xc}^+$ that demarcates the active and  passive subspaces that can be synchronized 
collapses to a single wavenumber, corresponding to the critical wavenength $l_{xc}^+=130$. It is evident that the components that are sustained in  RNL (with  $k_x^+$ much smaller than $k_{xc}^+$)
 constitute the main sources of energy transfer from the mean flow. 

The gravest streamwise components of the passive subspace $\u_>$, which are  $k_x=2\alpha$ for $R600$, $k_x=5\alpha$ for $R1500$ and $k_x=6\alpha$ for $R2250$, are located in a region of the spectrum where the linear production $\Pi_0$ has been reduced significantly and has been just overtaken by the $\Pi_{k_x}$ fluxes as the primary source of energy input.
An increase in $k^+_x$ shifts the balance of the components in the passive subspace to an equilibrium between $\Pi_{k_x}$ and $\epsilon_{k_x}$ at every
time instant  (cf. the black and blue lines shown in Fig. \ref{fig:Ekx_Uw}b), indicating  that dissipative dynamics eventually fully govern this part of the spectrum.

It is remarkable that scales as large as $130$ wall units that receive energy both from the mean flow and from nonlinear scattering, cf.   Fig. \ref{fig:Ekx_Uw}a, are slave to the larger scale flow.
This implies that the larger scale flow organizes the nonlinear transfers to the smaller scales and the backscattering from the smaller to the larger scales within the passive subspace is not
able to dictate the evolution. 

 \section{Discussion}
 
In the works of \cite{Alves-etal-2020} and \cite{Thiesset-etal-2014}, the small scale component that is hypothesized to describe the stochastic turbulence was found to be influenced by the time-dependent large-scale coherent structures of the flow. \cite{Alves-etal-2020} also identified a range of length scales where the components of the energetics are maintained in an equilibrium  between dissipation and the nonlinear fluxes.  
 We can therefore assume that the passive subspace $\u_>$ defined in this work is related to the stochastic component and  the highly influenced statistics found in those studies are a consequence of the synchronization imposed to the passive subspace from the coherent motions of the active subspace identified here
 with $\u_<$.

The feasibility of synchronization reveals a significantly reduced number of active components in the flow which are found to exceed a streamwise length of $l_x^+=130$ wall units in at least three different Reynolds numbers. We have verified the same 
wall unit threshold in all the Couette channels used for this study. The scaling with wall units suggests that the subspaces describe states that emerge near the wall.
 The cutoff length scale coincides with the peaks found by \cite{Nikitin-2018} in the spectra of $v$ and $w$ at $y^+=14$ for the Lyapunov vector of the total flow at $R_{\tau}=391$. Interestingly, once we restrict the streamwise length scale of the subspaces below this peak, the subspaces are stabilized.   
This finding could imply that an inverse cascade originates from $v$ and $w$ components of the cuttoff length scale.
Even though the experiments were performed in low Reynolds numbers, we assume that the wavelength cutoff of the small subspace will be retained and remain relevant even in high $R_{\tau}$ experiments. It is expected that synchronization times will rise as more wavenumbers are needed to resolve a simulation, since the increasing density of wavenumbers will facilitate the existence of components closer to the cutoff.

Another comparison can be made with the length scales attributed to the Kolmogorov dissipation range.
Translating $l_x^+=130$ into Kolmogorov lengths, defined as $\eta = (1/ (R^3 \bar{\epsilon}) )^{(1/4)} $ ( for $R1500$ the time-averaged dissipation $\bar{\epsilon} = 5.95 D_C$ is $5.95$ times the laminar dissipation rate $D_C$ of Couette flow), we find that this length scale equals roughly  $500\eta$ and is well above the assumed dissipation subrange cutoff of $l \approx 60 \eta$, showing that if those scales are relevant in high Reynolds number simulations they will also describe a portion of the motions that belong to the inertial subrange of the spectrum.
The synchronization experiments in isotropic turbulence \citep{Yoshida-etal-2005,Lalescu-etal-2013,Di-Leoni-etal-2020,Vela-Martin-2021} have recovered a cutoff wavenumber for synchronisation between $k \eta = 0.15-0.2$. When we scale the growth rates and wavenumbers in the same units we find that the value of our synchronization threshold is above $k_x \eta \approx 0.015$ which
implies that synchronization in those experiments occurs at smaller wavelengths than what we have obtained in this work.
 Comparisons between homogeneous isotropic and rotating 3D turbulence have shown that the presence of large-scale structure in rotating turbulence reduces significantly the amount of neccessary input to achieve synchronization \citep{Di-Leoni-etal-2020}. Such large scale structures are also prevalent in the wall-bounded flows studied in this work, which we consider the main cause for the difference found on the lengthscale of the synchronizable scales.

\section{Conclusions}

We have employed a streamwise Fourier decomposition of turbulent channel flow  to demonstrate that synchronization occurs in the Fourier subspaces comprised by the Fourier modes with streamwise wavelengths shorter  than $l_x^+=130$. 
The existence of this threshold was verified in a series of experiments at  low Reynolds numbers and its' connection with the top Lyapunov vector implies that the critical wavelength scaling will hold for flows where this vector is concentrated in the buffer layer region (which has been shown by \cite{Nikitin-2018} to apply for the top Lyapunov vector in flows up to $R_{\tau}=586$), where the large scale structure is comprised by rolls and streaks.

The active subspace spans only a fraction of the streamwise spectrum but the energy contained in these components
is a significant portion of the total kinetic energy,  as  was shown to be the case 
also  in isotropic turbulence \citep{Yoshida-etal-2005, Di-Leoni-etal-2020}. 
The components of the active subspace are mainly responsible for the energy extraction from the mean flow and have been shown to be piecewise unsynchronizable when they belong to the RNL active subspace, suggesting that RNL captures the source of chaotic dynamics in turbulent flows.

The present paper shows that the state of the active subspace uniquely determines the passive subspace.
The passive subspace can be recovered exactly once the mean flow and the top streamwise harmonics that comprise the large scale structure have been obtained,  and therefore implies that the active degrees of freedom in the dynamics are significantly reduced. 
Evidence pointing to synchronization phenomena has also been found in other inhomogeneous flow configurations, where the statistics and structure functions of the small scale stochastic fluctuations are strongly influenced by the large scale coherent component of the flow
\citep{Thiesset-etal-2014,Alves-etal-2020}.

Finally, from the fact that any finite amplitude deviations of the passive subspace decay 
we conclude that those scales belong to the inertial manifold of the system.
Inertial manifolds define parts of the dynamics that govern the finer details of the attractor and can be neglected when we are interested in a more qualitative view of the dynamics.
However, it is still necessary to account for this subspace in the dynamics in order to find an accurate solution for the active subspace. 
Further research on the relation between the two subspaces could determine if it is possible to formulate an expression for the passive subspace as a function of the active.

 \bibliographystyle{jfm}

\end{document}